\documentclass[aps,twocolumn, pra]{revtex4}
\usepackage{graphicx}
\usepackage{epsfig}
\usepackage{newlfont}
\usepackage{amssymb}
\usepackage{amsfonts}

\usepackage{amsmath}
\usepackage{mathbbol}
\usepackage{graphicx}
\usepackage{bm}

\begin{document}
\title{Atom Counting in Expanding Ultracold Clouds}

\author{Sibylle Braungardt\(^1\), Mirta Rodr\' iguez\(^2\), Aditi Sen(De)\(^3\), Ujjwal Sen\(^3\),and Maciej Lewenstein\(^{1,4}\)}
\affiliation{\(^1\)ICFO-Institut de Ci\`encies Fot\`oniques,
Mediterranean Technology Park, 08860 Castelldefels (Barcelona),
Spain\\ \(^2\) Instituto de Estructura de la Materia, CSIC,
Serrano 121, 28006 Madrid, Spain\\\(^3\) Harish-Chandra Research
Institute, Chhatnag Road, Jhunsi, Allahabad 211 019, India\\
\(^4\) ICREA - Instituci\'{o} Catalana de Recerca i Estudis
Avan\c{c}ats, Passeig Lluis Companys 23, E-08010 Barcelona, Spain}

\begin{abstract}
We study the counting statistics of ultracold bosonic atoms that
are released from an optical lattice. We show that the counting
probability distribution of the atoms collected at a detector
located far away from the optical lattice can be used as a method
to infer the properties of the initially trapped states. We
consider initial superfluid and insulating states with different
occupation patterns. We analyze how the correlations between the
initially trapped modes that develop during the expansion in the
gravitational field are reflected in the counting distribution. We
find that for detectors that are large compared to the size of the
expanded wave function, the long-range correlations of the initial
states can be distinguished by observing the counting statistics.
We consider counting at one detector, as well as the joint
probability distribution of counting particles at two detectors.
We show that using detectors that are small compared to the size
of the expanded wave function, insulating states with different
occupation patterns, as well as supersolid states with different
density distributions can be distinguished.
 \end{abstract}
\maketitle

\def\com#1{{\tt [\hskip.5cm #1 \hskip.5cm ]}}

\def\bra#1{\langle#1|} \def\ket#1{|#1\rangle}
\def \av#1{\langle #1\rangle}
\section{Introduction}
Experiments with ultracold particles trapped in optical lattices
aim towards the engineering of exotic many-body quantum states
\cite{adp}. Recently, the trapping and cooling of dipolar gases
have attracted much attention \cite{Lahaye2009}. The dipole
moments induce long-range interactions between the particles, and
new phases appear \cite{metastable}. In the strongly correlated
regime, it has been shown that there are many quasi degenerate
metastable insulating states with defined occupation patterns
\cite{Menotti2007,Trefzger2008,Capogrosso2010,Pollet2010}. These
metastable states could be used for the storage and processing of
quantum information in analogy to classical neural networks, where
the information is robustly encoded in the distributed stable
states of a complex system \cite{Dorner2003,Pons2007}. Another way
to induce long-range interactions between atoms trapped in an
optical lattice is via coupling to an external cavity mode. This
has just recently been achieved experimentally and a checkerboard
to a supersolid transition has been observed \cite{Baumann2010}.

 The detection of exotic
strongly correlated phases requires novel experimental techniques
that give access to high-order correlation functions. Proposals
for detection techniques typically make use of shot-noise
measurements \cite{shotnoise} or atom-light interfaces \cite{ali}.
Also, the counting statistics of atoms has been suggested as a
technique able to distinguish strongly correlated
\cite{amader_counting,demler} and fermionic \cite{fermiones}
Hamiltonians, both at zero and finite temperature
\cite{braungardt2011}.
%epxerimentally counting on site available
The detection of single atoms trapped in the optical lattice has
become experimentally available
\cite{Gericke2008,Chen2009,Bakr2010,Sherson2010} only recently.
Most counting experiments
 are performed after switching off the trapping potential and letting the atoms
propagate in the gravitational field.
% mainly experiments of counting are done after expansion
The counting statistics of Rb atoms falling within a high-finesse
cavity has been reported in Ref. \cite{esslinger04}. Also,
fermionic and bosonic counting probability distributions have been
measured for metastable Helium atoms falling onto a microchannel
plate \cite{Aspect_science, Aspect_nature}.

The theoretical analysis of the counting process has so far mainly
been considered for atoms trapped
  in the lattice. Propagation in the gravitational field mixes the initial modes of the atoms,
  such that
the counting statistics in the lattice and after propagation are
not expected to be the same. In this paper, we study the role of
expansion in the counting process. We show that the mixing of the
initial modes during the expansion becomes evident in the counting
distribution when the detector is small compared to the size of
the expanded wave function. We illustrate the effect by analyzing
the counting statistics for bosons after time-of-flight expansion
from the lattice. We consider initial states with different
occupation patterns in the insulating regime and supersolid states
with different density distributions in the superfluid regime. We
calculate both the counting probabilities at a single detector and
the joint probabilities at two detectors as a function of the
horizontal distance between them. We show that a superfluid (SF)
and Mott insulator (MI) state can be readily distinguished by
their counting statistics. We further show that a suitable choice
of the detector geometry allows for the detection of different
occupation patterns in the insulating regime and different
supersolid states.

The paper is organized as follows. In Sec. \ref{s1} we review the
propagation of the atomic wave functions and the atom counting
formalism. In Sec. \ref{s2} we analyze the intensity of particles
arriving at the detector, which consists of auto-correlation terms
and crossed-correlations between the different expanded modes.
Depending on the size and geometry of the detector, the ratio
between the auto-correlations and the crossed-correlation terms
changes. In Sec. \ref{s3} we obtain closed expressions for the
counting distributions for expanded superfluid and insulating
bosonic states. We consider the counting statistics when using one
detector and the joint counting distribution at two detectors. In
Sec. \ref{res}, we show our results and compare the SF with MI
states and insulating states with different occupation patterns.

\section{Description of the System \label{s1}}
We consider neutral bosonic atoms trapped in an optical lattice.
The system can be described using the Bose-Hubbard model
\cite{bose_hubbard}, which includes the hopping of the particles
between neighbouring sites and the on-site two-body interactions.
At zero temperature, the two limiting cases of the phase diagram
are the SF state, where the hopping term dominates, and the MI
state, where local interactions are dominant. The field operator
$\Psi (\textbf{r},t)$ of the many-body system can be expanded into
the $N$ modes $a_i$
\begin{equation}
\Psi (\textbf{r},t)=\sum_i\phi_i(\textbf{r},t) a_i.\label{eq-psi}
\end{equation}
For atoms trapped in an optical lattice, $a_i$ describes the
destruction of a particle on site $i$. The corresponding initial
wave functions are Wannier functions which are Gaussian functions
centered at $\textbf{r}_i$
\begin{equation}
\phi_i(\textbf{r},t=0)=\frac{1}{(\pi\omega^2)^{3/4}}e^{-(\textbf{r}-\textbf{r}_i)^2/2\omega^2},\label{eq-phi}
\end{equation}
where the width $\omega$ is chosen such that the initial wave
functions at different sites $i$ do not overlap.

 The atoms are
released from the optical lattice and expand in the gravitational
field. At finite $t$, we can apply the single-particle expansion
\begin{equation}
 \phi(\textbf{r},t)=\int d \textbf{r}'
K(\textbf{r},\textbf{r}',t) \phi_i(\textbf{r}',0)
\end{equation}
where the propagator for the free expansion in the gravitational field reads \cite{propagator}
\begin{equation}
K(\textbf{r},\textbf{r}',t)=\left(\frac{m}{2\pi i\hbar t}
\right)^{3/2}e^{\frac{im(\textbf{r}-\textbf{r}')^2}{2\hbar
t}-\frac{imgt(z+z')}{2\hbar}-\frac{im^2g^2t^3}{24m\hbar}}.
\end{equation}
The full propagated wave function is then written as
\begin{eqnarray}
\phi_i(\textbf{r},t) =  \frac{e^{-\frac{im^2g^2t^3}{24m\hbar}}}{\pi^{3/4}(i
\omega_t+\omega)^{3/2}}e^{-\frac{(\textbf{r}-\textbf{r}_i)^2}{2(\omega_t^2+\omega^2)}}e^{-i\frac{(\textbf{r}-\textbf{r}_i)^2\omega_t}{2\omega(\omega_t^2+\omega^2)}}, \label{eq:fi}
\end{eqnarray}
where and $\textbf{r}_t=\textbf{r}+\textbf{z}_t$, with
$\textbf{z}_t=(0,0,gt^2/2)$ and we have used that
$|\textbf{r}_t-\textbf{r}_i|\gg \omega$. Note that in the limit of
$\omega_t \gg \omega$, the expanded wave function is, up to a
phase factor, a Gaussian function centered around $\textbf{z}_t$
with a width $\omega_t=\hbar t/(m \omega)$.

%In the context of particle counting, we are mainly interested in
%the number operator at a point $\textbf{r}$ at time $t$. Using eq.
%(\ref{phi}), we can calculate the ,
%\begin{equation}
%\Psi^\dagger (\textbf{r},t) \Psi
%(\textbf{r},t)=\sum_{ij}\phi_i^*(\textbf{r},t)\phi_j(\textbf{r},t)b_i^\dag
%b_j.\label{n_Psi}
%\end{equation}

\subsection{Atom counting}
We describe a counting process in which the probability $p(m)$ of
counting $m$ particles within a time interval $\tau$ is measured
at a detector located at a distance $z_0$ from the lattice. The
probability of detecting $m$ particles can be expressed as
\cite{Glauber,Cahill_Glauber}
\begin{equation}\label{eq-p}
p(m)=\frac{(-1)^m}{m!}\frac{d^m}{d\lambda^m}\mathcal{Q}\Big|_{\lambda=1},
\end{equation}
where the generating function $\mathcal{Q}(\lambda)$  is given by
the expectation value of a normally ordered exponential of the
intensity $\mathcal{I}$,
\begin{equation}
\mathcal{Q}(\lambda)=\mbox{Tr}(\rho:e^{-\lambda\mathcal{I}}:).\label{eq-Q}
\end{equation}
For photons, the intensity is proportional to an integral over the
product of the negative-frequency part and the positive-frequency
part of the field. The normal ordering $: ... :$ reflects the
detection mechanism, in which the photons are absorbed at the
detector, typically a photo multiplier or an avalanche photodiode.
For the detection of atoms using microchannel plates, the
detection process can be treated in an analogous way. %CONFIRM IF
%THIS IS TRUE FOR ABSORPTION IMAGING.

Since typically not all the particles are counted, the intensity
depends on the efficiency $\epsilon$ of the detector and the
detection time $\tau$. When the dynamics of the measurement are
fast in comparison to the dynamics of the system, the intensity is
proportional to the factor $\kappa\equiv1-e^{-\epsilon\tau}$.
 For typical experimental situations, the dynamics of the system
 are
determined by the expansion of the atomic cloud in the
gravitational field, given by $\omega_t$ and the intensity can be
described by the integral over the detector volume $\Omega$ of the
positive-frequency and negative-frequency parts of the quantum
fields describing the particles to be counted, multiplied by the
efficiency factor $\kappa$ \cite{grochmalicki},
\begin{eqnarray}
\mathcal{I}=\kappa \int_{\Omega} d\textbf{r}\Psi^\dagger
(\textbf{r},t_d) \Psi (\textbf{r},t_d)\label{intensity},
\end{eqnarray}
where $t_d$ denotes the time at which the instantaneous
measurement is performed.

% we calculate p(m) mean and variance
%Important magnitudes we look at are the full number distribution, the mean and the variance,

%\subsection{Counting with more than one
%detector} \label{Sec_two_det}
The formalism described above is easily generalized to the case of
detection with multiple detectors \cite{Arecchi1966}. For
detection with $M$ detectors, the generating function reads
\begin{equation}
\mathcal{Q}_M(\lambda_1,\lambda_2,..,\lambda_M)=\mbox{Tr}(\rho:e^{-\sum_i\lambda_i\mathcal{I}_i}:),
\end{equation}
where the single detector intensity $\mathcal{I}_i$ for each of
the detectors is given by eq. (\ref{intensity}). For a
configuration with two detectors, the joint probability
distribution of counting $m$ atoms at detector $1$ and $n$ atoms
at detector $2$ is given by
\begin{equation}
p(m,n)=\frac{(-1)^{m+n}}{m!n!}\frac{d^{m+n}}{d\lambda_1^md\lambda_2^n}\mathcal{Q}_2\Big|_{\lambda_1=1,\lambda_2=1}.\label{joined_p}
\end{equation}
We study the correlations $\hbox{corr}(m,n)$ between the counting
events detected at each detector by observing the ratio between
the covariance and the single detector variances,
\begin{equation}
\hbox{corr}(m,n)=\frac{\hbox{cov}(m,n)}{\sigma^2(m)\sigma^2(n)},\label{eq-correlations}
\end{equation}
where $\hbox{cov}(m,n)=\sum_{m,n} m n p(m,n)-\bar{m}\bar{n}$,
$\bar{m}$ denotes the mean and $\sigma^2(m)$ the variance of
$p(m)$.
%var(m)=var(n)=\nonumber\\\sum_{m,n} m^2
%p(m,n)-\bar{m}^2=\sum_{m,n} n^2
%p(m,n)-\bar{n}^2\\
%\bar{m}=\bar{n}=\sum_{m,n} m p(m,n)=\sum_{m,n} n p(m,n)
%\end{eqnarray}

\section{Detection of expanding atoms \label{s2}}
Let us now discuss the counting process for the detection of atoms
expanding in the gravitational field. We consider a cubic detector
located at a distance $z_0$ from the lattice center with edge
lengths $\Delta_x, \Delta_y, \Delta_z$ . For simplicity, all
through this paper we consider $t_d=\sqrt{2 z_0/g}$ which is the
time when the center of the cloud arrives at the detector. The
intensity $\mathcal{I}$ of atoms registered at the detector
defined in eq. (\ref{intensity}) is thus determined by the
expanded field operator of the atoms at the time $t_d$ of
detection, $\Psi(z_0,t_d)$. Using eqs. (\ref{eq-psi}) and
(\ref{intensity}), the intensity $\mathcal{I}$ takes the form
\begin{equation}
\mathcal{I}=\sum_{ij}A_{ij} a_i^\dag a_j,
\end{equation}
where
\begin{equation}
A_{ij}(z_0,\Omega,\kappa)=\kappa\int_{\Omega}
d\textbf{r}\phi_i^*(z_0,t_d)\phi_j(z_0,t_d).\label{A}
\end{equation}
The elements of the correlation matrix $A_{ij}$ defined in eq.
(\ref{A}) describe the interference and autocorrelation terms
between different modes registered at the detector. The diagonal
terms represent the on-site correlations, whereas the off-diagonal
terms represent the crossed-correlations between single particle
modes initially located at different sites with distance $|i-j|$.

Before studying the full counting distribution, let us consider
the correlations given by the matrix elements $A_{ij}$. Using eq.
(\ref{eq:fi}) and assuming $\omega_{t_d}\gg \omega$, the
autocorrelation elements are given by
\begin{equation}
A_{ii}=\kappa \int_{\Omega} d\textbf{r}\frac{1}{\pi^{3/2}
\omega_{t_d}^3}e^{-\frac{(\textbf{r}-\textbf{r}_i)^2}{\omega_{t_d}^2}}.\label{eq-diag-A}
\end{equation}
For expanded wave functions at $\textbf{r} \gg \textbf{r}_i$, the
autocorrelations become all equal and independent of the original lattice site
$i$. The crossed-correlations are given by
\begin{equation}
A_{ij}=\kappa \int_{\Omega} d\textbf{r} \frac{1}{\pi^{3/2}
\omega_t^3}e^{-\frac{(\textbf{r}-\textbf{r}_i)^2}{\omega_t^2}}e^{-i\frac{\textbf{r}(\textbf{r}_i-\textbf{r}_j)}{\omega\omega_t}}\label{eq-off-diag-A}
\end{equation}
The ratio between the crossed correlations and the
auto-correlations depend crucially on the geometry of the
detector.

\begin{figure}
\epsfig{file= 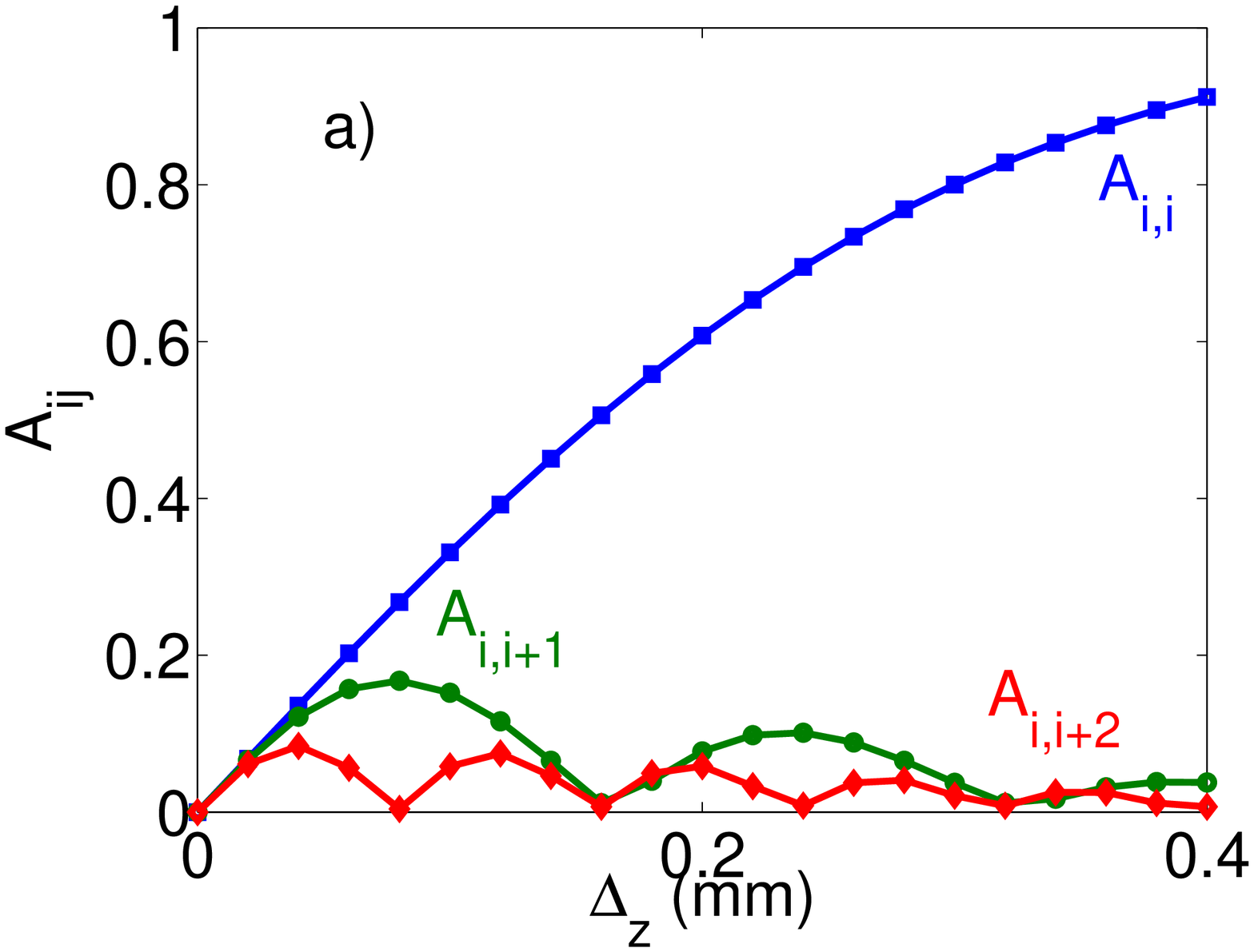,width=0.9\linewidth} \epsfig{file=
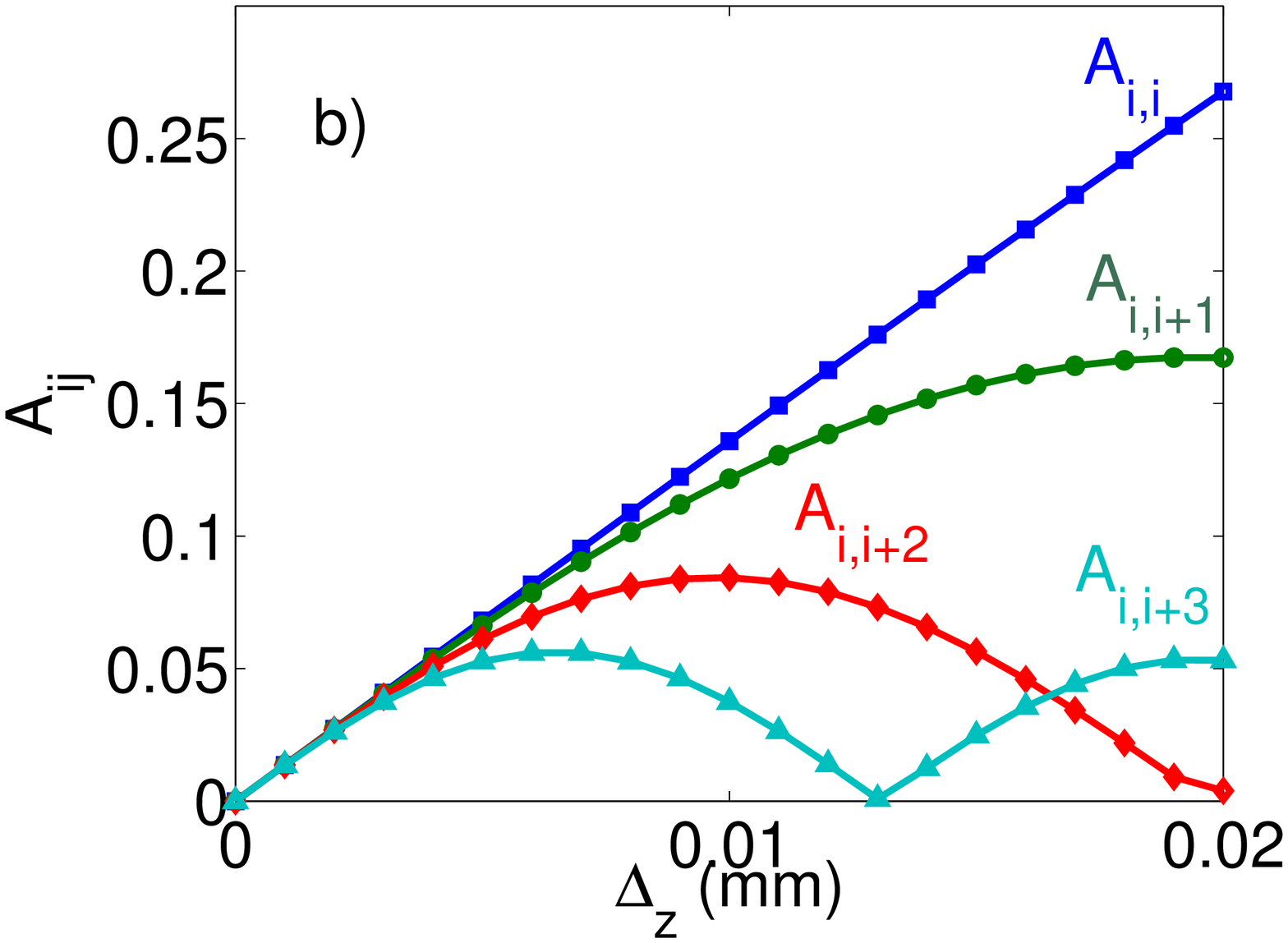,width=0.9\linewidth}\caption{The ratio between the
on-site-correlations and interference terms depend on the size of
the detector. a) Wide detector limit $\Delta_z \simeq \omega_t$. b) Narrow
detector limit $\Delta_z\ll\omega_t$. We plot $A_{ii}$ (blue
squares) $A_{i,i+1}$ (green circles), $A_{i,i+2}$ (red diamonds),
 and $A_{i,i+3}$ (light blue triangles). Parameters used: $z_0=1$cm,
$\Delta_x=\Delta_y=1 $cm and $\kappa=1$ } \label{fig-Elements-A}
\end{figure}

 In Fig. \ref{fig-Elements-A}, we show the
on-site correlations eq. (\ref{eq-diag-A}) and the interference
terms eq. (\ref{eq-off-diag-A}) in function of the size of the
detector. We consider a one dimensional array in $z$-direction and
plot the correlations at the location of the detector at
$(0,0,z_0)$. We consider a fixed detector size in the x-y-plane,
$\Delta=\Delta_x=\Delta_y$ and vary its width $\Delta_z$.
Depending on the size of the detector, the whole cloud or a
fraction of it is registered. For $z_0=1$ cm, the size of the
expanded single-particle wave function at the detector is
$\omega_{t_d}=0.8$ mm.
In Fig. \ref{fig-Elements-A} a, we show that for detectors of size
$\Delta_z > 0.2$ mm, the interference terms are negligible. This
is easily understood from eq. (\ref{eq-diag-A}), as the
auto-correlations are given by an integral over the detector
volume around the center of a Gaussian function. For detectors
that are large compared to the size of the cloud, the on-site
correlations approach unity. In contrast, the interference terms
eq. (\ref{eq-off-diag-A}) are given by the integral over a
Gaussian function multiplied by a highly oscillating phase, such
that they approach zero as the size of the detector increases.
Thus, for on-site counting and for detectors which are larger than
the size of the cloud, the crossed correlations disappear
$A_{ij}\simeq 0$ for $i \neq j$, while the auto-correlations
approach $A_{ii} \simeq \kappa$.

As we show below, the detection of the auto-correlations between
different modes is sufficient to distinguish the long-range
correlations in the system. In particular, we show that a MI state
can be distinguished from a SF.

On the contrary, as the auto-correlation terms for different sites
are equal, distinguishing states with different occupation
patterns cannot be achieved in this limit. Fig.
\ref{fig-Elements-A} b shows that for small detector sizes, the
interference terms are of the order of the on-site correlations.
We will show that in this limit, different occupation patterns are
distinguishable from the counting distribution.
\section{Atom Counting Statistics \label{s3}}

Let us now consider the counting distributions measured at the
detector after the expansion for different initial states of the
system of atoms trapped in the lattice.

\subsection{Superfluid state}
First, let us focus on a SF state, ground state of the
Bose-Hubbard model for very shallow lattices. We derive the
counting distribution using the Gutzwiller ansatz
\cite{gutzwiller} for the wave function which assumes that it is a
product of on-site coherent states. The initial state of the atoms
in the lattice with \(N\) sites then reads:
\begin{equation}
\ket{\psi}=\prod_i^N\ket{\alpha_i}_i, \label{sf}
\end{equation}
where $\ket{\alpha_i}_i$ is the coherent state on site $i$,
\begin{equation}
\ket{\alpha_i}_i=e^{-|\alpha_i|^2/2}\sum_{n=0}^\infty
\frac{\alpha_i^n}{\sqrt{n!}}\ket{n}_i\,\label{eq:cs}
\end{equation}
and $\ket{n}_i=(a_i)^n\ket{0}$ is a Fock state with $n$ particles.
Note that $\ket{\psi}$ is an eigenstate of the annihilation
operator $\Psi(\textbf{r},t)$ of the expanded atoms,
\begin{eqnarray}
\Psi(\textbf{r},t)\ket{\psi}=
% \sum_i\phi_i(\textbf{r},t)  b_i\prod_k\ket{\alpha_k}\nonumber\\
\sum_i\phi_i(\textbf{r},t)\alpha_i\ket{\psi},
\end{eqnarray}
where $\phi_i$ is given by eq. (\ref{eq-phi}). The state
$\ket{\psi}$ is thus an eigenstate of the expanded field operator
$\Psi(\textbf{r},t)$ and we can write the generating function as
$\mathcal{Q}(\lambda)=e^{-\lambda\sum_{ij}\alpha_i^*\alpha_j
A_{ij}}$. Using eq. (\ref{eq-p}) the counting distribution $p(m)$
reads
\begin{eqnarray}
\label{eq-p_SF} && p(m)=\frac{\left(
\sum_{ij}\alpha_i^*\alpha_jA_{ij}\right)^m}{m!}e^{-
\sum_{ij}\alpha_i^*\alpha_jA_{ij}},
\end{eqnarray}
where $A_{ij}$ is given by eq. (\ref{A}).

%%%%%%%%%%%%%
For a homogeneous superfluid with equal mean number of particles
per sites, $\alpha_i=\alpha$ for all $i$, and in the limit of big
detectors where the diagonal elements of the matrix $A_{ij}$ are
much bigger than the off-diagonal elements, the counting
distribution of the SF simplifies to
\begin{equation}
p(m)=\frac{(N|\alpha|^2A_d)^m}{m!}e^{-N|\alpha|^2A_{d}},\label{eq-p-sf}
\end{equation}
which corresponds to a Poissonian distribution with mean (and thus
also variance) $\bar{m}=\sigma^2(m)=|\alpha| A_d$.
\subsection{Mott Insulator state}
Let us now consider the Mott insulating regime. We first study a
Mott insulator state with one particle per site,
$\ket{\psi}=\ket{11..11}$. In this case, the generating function
eq. (\ref{eq-Q}) reads
\begin{eqnarray}
\mathcal{Q}(\lambda)&=&\bra{11..11}:e^{-\lambda\kappa\int_\Omega
d\textbf{r}\Psi^\dag(\textbf{r},t_d)\Psi(\textbf{r},t_d)}:\ket{11..11}
\nonumber\\
&=&1-\lambda\sum_iA_{ii}+\lambda^2\sum_{i<j}(A_{ii}A_{jj}+|A_{ij}|^2)-...
.\label{eq-mottA}
\end{eqnarray}
 We can rewrite eq.
(\ref{eq-mottA}) using the minors of the matrix $A$,
\begin{eqnarray}
\mathcal{Q}(\lambda)=1+\sum_{k=1}^N (-1)^k\lambda^k
\hbox{M}_{+}(A,k),\label{Q_Mottbf}
\end{eqnarray}
where M$_{+}(A,m)$ denotes the permanent
perm$(A)=\sum_{\sigma\epsilon S_n}\Pi_{i=1}^nA_{i,\sigma(i)}$ of
the corner blocks of size $m$ of the matrix $A$. Note that
M$_{+}(A,k)$ is closely related to the principal minors of the
matrix, which are defined as the determinant of the respective
block matrices. The counting distribution $p(m)$ can then be
calculated using Eqs.~(\ref{eq-p}) and (\ref{Q_Mottbf}).

As was outlined above, in typical experimental situations the
detector is far away from the lattice and much bigger than the
cloud, such that the off-diagonal elements of $A_{ij}$ are
negligible and the diagonal elements $A_{ii}$ are equal for all
$i$. In this case the generating function $\mathcal{Q}$ for the
Mott insulator state with unit filling is given by
\begin{equation}
\mathcal{Q}(\lambda)=\sum_{k=0}^N\binom{N}{k}(-\lambda
A_{d})^k=(1-A_d\lambda)^N,\label{MI_high_eff}
\end{equation}
where $A_{d}$ denotes any of the (equal) diagonal elements.
 The
counting distribution $p(m)$ is then given by
\begin{eqnarray}
p(m)=\binom{N}{m}A_{d}^m(1-A_{d})^{N-m}
\end{eqnarray}
This corresponds to the distribution of a fock state. The mean
$\bar{m}$ and variance $\sigma^2(m)$ of the distribution are given
by
\begin{equation}
\bar{m}=N A_d,\,\,\,\,\,\, \sigma^2(m)=N A_{d}(1-A_d)
\end{equation}

Let us now consider the different occupation patterns that arise
in the strongly correlated regime. In particular, we focus on such
states where at most one particle occupies each site. The
generating function is then calculated by eq. (\ref{Q_Mottbf}),
with a correlation matrix $A'$, composed of the elements of the
correlation matrix $A$ in eq. (\ref{A}) multiplied by the
occupation numbers $n_i$ and $n_j$ of the involved sites,
\begin{equation}
A'=n_i n_j A_{ij}.\label{A-patterns}
\end{equation}

Finally, let us consider a symmetric superposition of all possible
states with filling factor $N_p/N_s$, where $N_p$ is the number of
particles, $N_s$ denotes the number of sites and $N_p \leq N_s$,
the generating function reads
\begin{eqnarray}
\mathcal{Q}=1+\sum_m(-1)^m\lambda^m\mathcal{F}_{MI} (A,m,N_p,N_s)
,\label{eq-Q-Mott-general}
\end{eqnarray}
where
\begin{eqnarray}
\mathcal{F}_{MI} (A,m,N_p,N_s)=\frac{\binom{N_s-m}{N_p-m}}{\binom{N_s}{N_p}}M^+(A,m)\nonumber\\
+\frac{\binom{N_s-2m}{N_p-m}}{\binom{N_s}{N_p}}2^m\mathcal{K}(m),\label{eq:FMIl}
\end{eqnarray}
where $\mathcal{K}(m)$ is defined as the $m$fold product over the
sum with non-repeated indices of the real part of $A_{ij}$,
$\sum_{i<j}\hbox{Re}(A_{ij})$ . For $m=2$, e.g.
$M_{+}=\sum_{i<j}(A_{ii}A_{jj}+|A_{ij}|^2)$ and
$\mathcal{K}(m)=\hbox{Re}(A_{ij})\hbox{Re}(A_{kl})$ with $k,l \neq
i,j$.

\subsection{Counting at two detectors}
In this section, we consider the detection of the MI and SF state
using two detectors and study the correlations between the
counting events.

For the MI state, the joint counting distribution $p(m,n)$ of
counting $m$ particles at one detector and $n$ particles at the
other is
 given by eq. (\ref{joined_p}),
 where the generating function for two detectors is given by
 \begin{equation}
 \mathcal{Q}_{2}= \sum_{k=1}^N (-1)^k
M^+(\lambda_1A^{(1)}+\lambda_2A^{(2)},k).\label{eq-MI-joint}\end{equation}
For detectors that are located symmetrically with respect to the
origin in the $x$-$y$-plane, in typical experimental situations
the off-diagonal elements of $A_{ij}$ are negligible (see fig.
\ref{fig-Elements-A}), and the diagonal elements $A_d$ are all
equal for both detectors, $A^{(1)}_d=A^{(2)}_d=A_d$. The
generating function thus simplifies to
\begin{eqnarray}
\mathcal{Q}_2(\lambda_1,\lambda_2)=\sum_{k=0}^N\binom{N}{k}(-A_{d})^k(\lambda_1+\lambda_2)^k\nonumber\\
=(1-A_d(\lambda_1+\lambda_2))^N,\label{MI_high_eff}
\end{eqnarray}
and the counting distribution is given by
\begin{eqnarray}
& p(m,n)=(-1)^{n+m}(1-2A_d)^{Np-m-n} \times \nonumber \\&
(-A_d)^{m+n }\frac{Np!}{m!n!(Np-m-n)!}
\end{eqnarray}

For the SF state, the joint counting distribution $p_{SF}(m,n)$ is
the product of the two single detector distributions $p_1(m)$ and
$p_2(n)$ given by eq. (\ref{eq-p_SF}). The counting events at the
two detectors are thus not correlated.
\section{Results \label{res}}
\subsection{Mott Insulator and Superfluid state}
\begin{figure}[t]
\epsfig{file=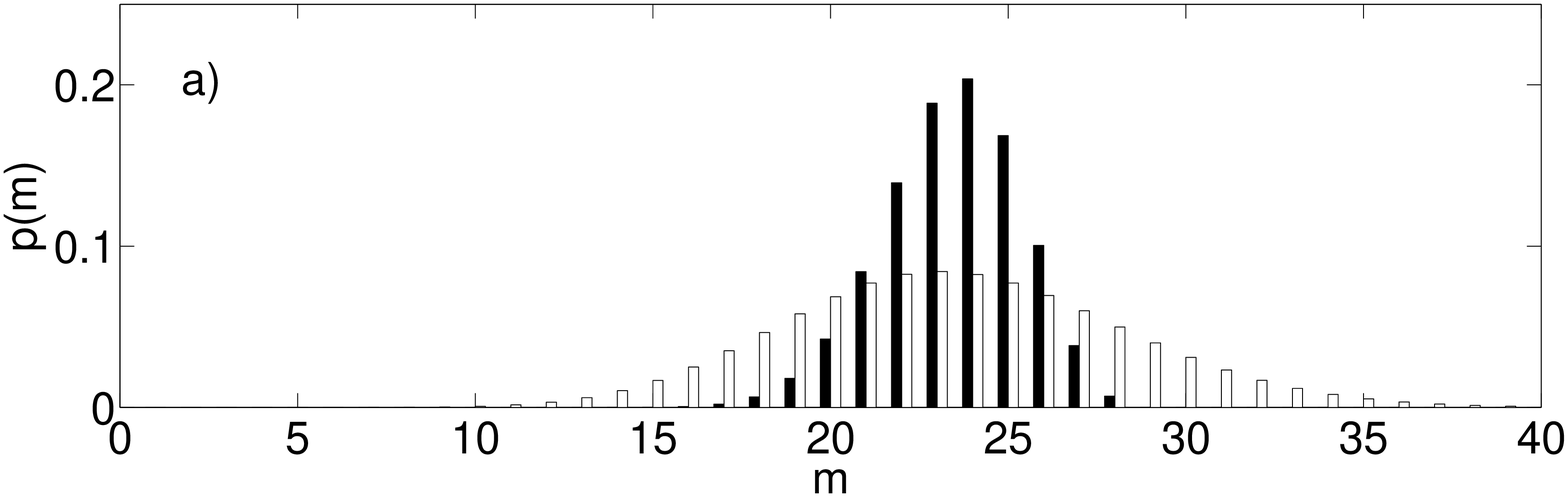,width=0.9\columnwidth}\\
\epsfig{file=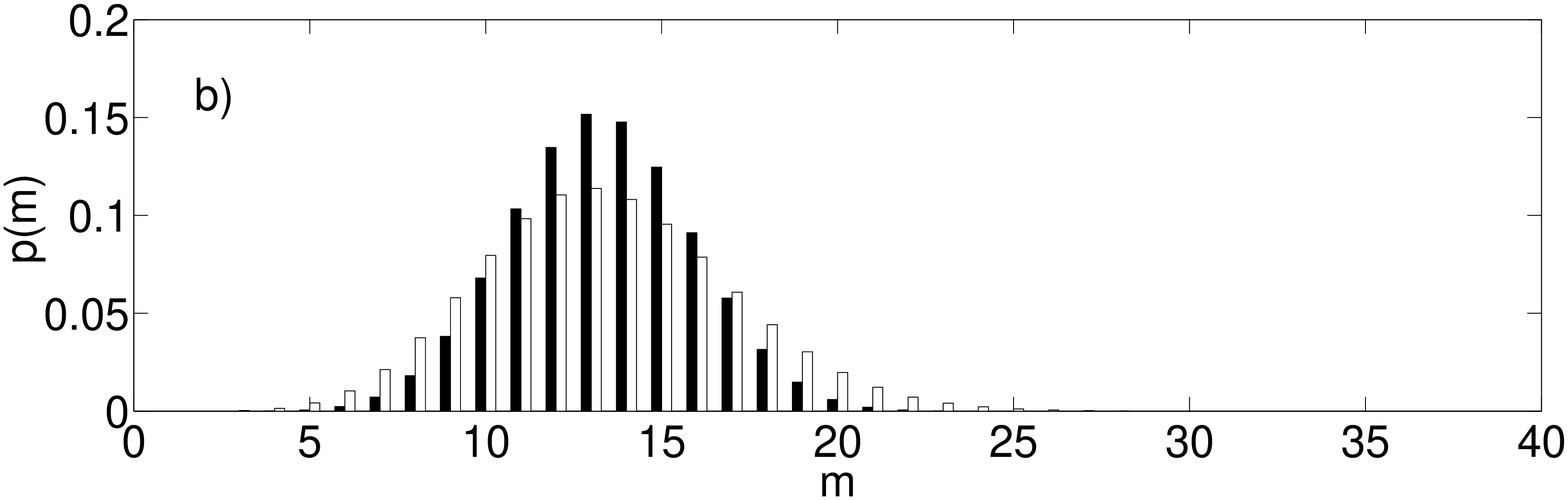,width=0.9\columnwidth}\\
\epsfig{file=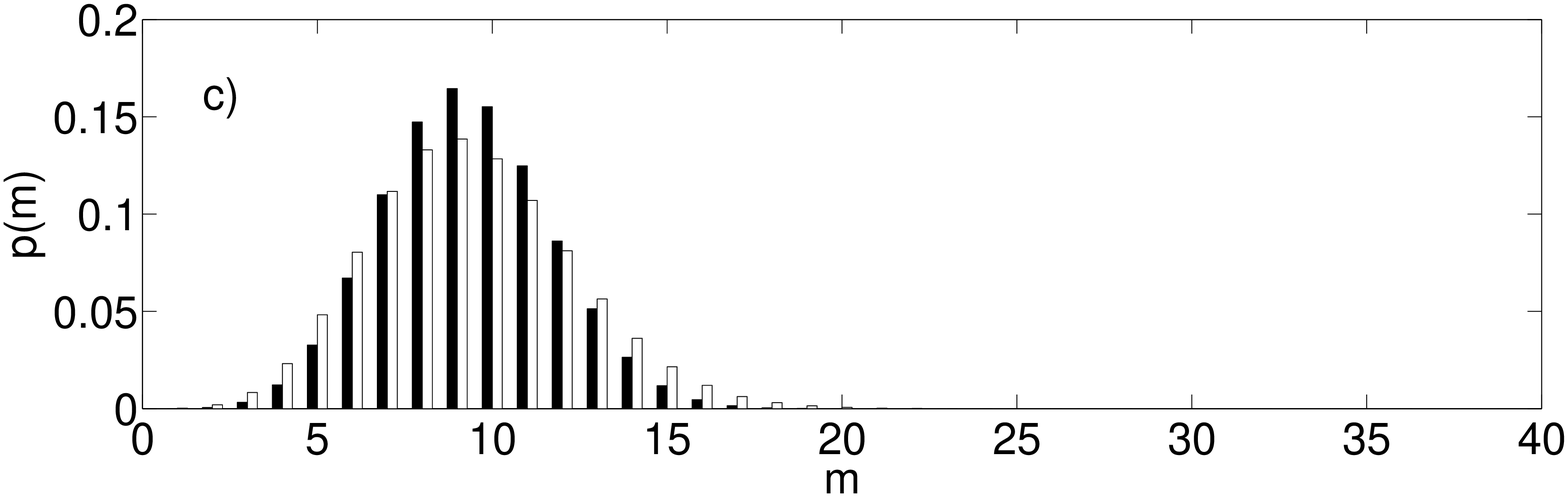,width=0.9\columnwidth}%\\\vspace{0.2cm}
\caption{MI vs. SF as a function of distance from detector.
Probability distribution for MI (black bars) and superfluid (white
bars) states in a 3x3x3 lattice. $\Delta_x=\Delta_y=2$ mm,
$\Delta_z=2$ cm, $\kappa=1$. a) $z_0=1$ cm, b) $z_0=3$ cm, c)
$z_0=5$ cm} \label{fig:distributions_mi_sf}
\end{figure}

\begin{figure}[t]
\epsfig{file=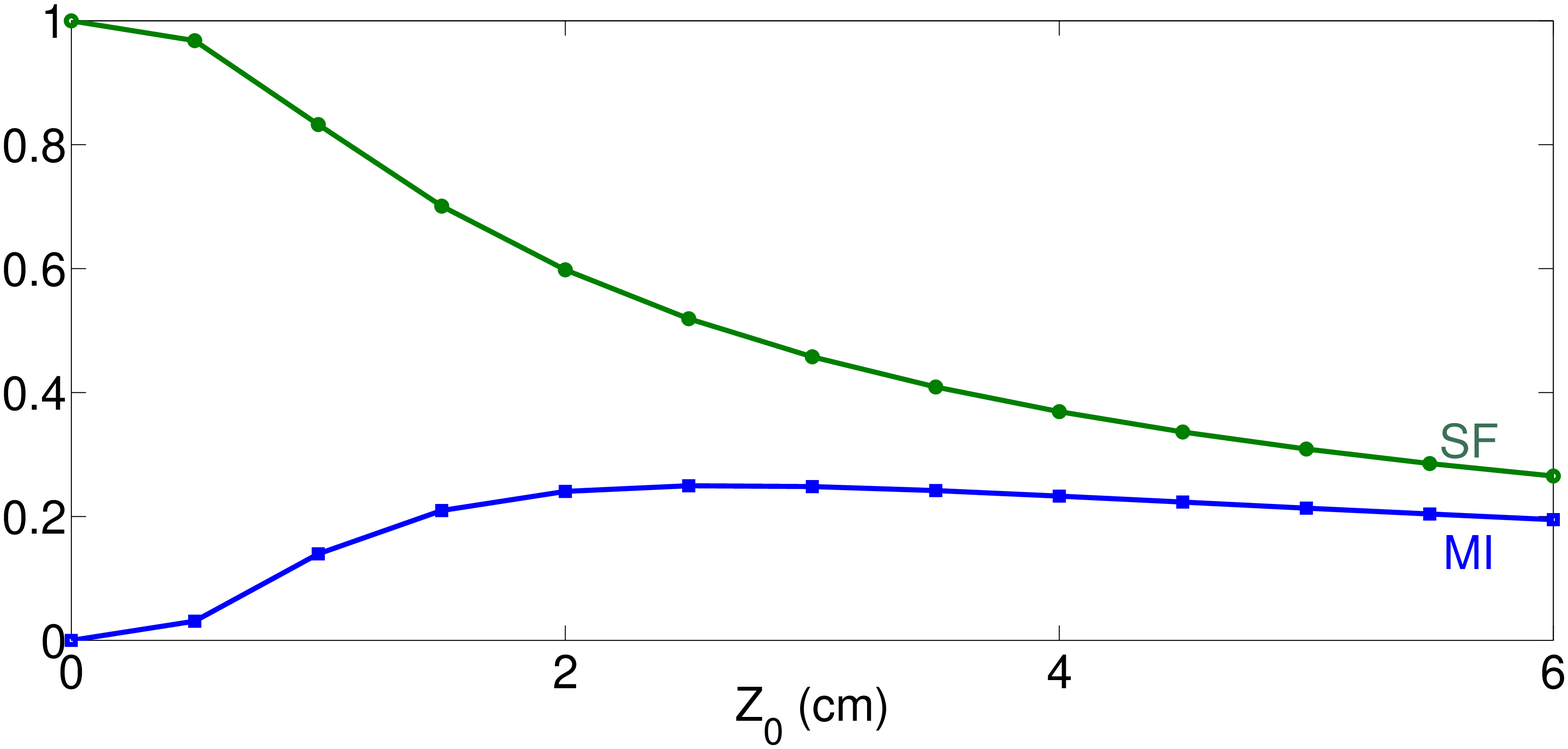,width=0.99\columnwidth}%\\\vspace{0.2cm}
\caption{$\sigma^2(m)/N$ of the counting distribution for the MI
(blue squares) and SF (green circles) state with respect to the
distance from the detector $z_0$. $\Delta_x=\Delta_y=2$ mm,
$\Delta_z=2$ cm, $\kappa=1$.} \label{fig:mi-sf-mean-var}
\end{figure}

We consider the counting distributions of a SF and a MI state of
bosons with the same average number of particles released from a
three dimensional optical lattice. We assume the limit of a large
detector, where the counting distribution is determined by the
on-site correlation terms. In Fig. \ref{fig:distributions_mi_sf},
we plot the counting distributions for a SF and a MI state at
different distances between the detector and the lattice. With
increasing distance from the detector, a smaller fraction of the
expanded wave function is registered. The difference between the
MI and the SF becomes less visible, and the mean of the counting
distribution decreases.
 In Fig. \ref{fig:mi-sf-mean-var}, we
plot the mean and the variance of the counting distributions, both
normalized by dividing by $N$, for the superfluid and the Mott
insulator state for a detector with fixed size at different
distances $z_0$ from the lattice.

\subsection{Mott Insulator and Superfluid state with two detectors}
\begin{figure}
\centering
\begin{tabular}{ccc}
\epsfig{file= 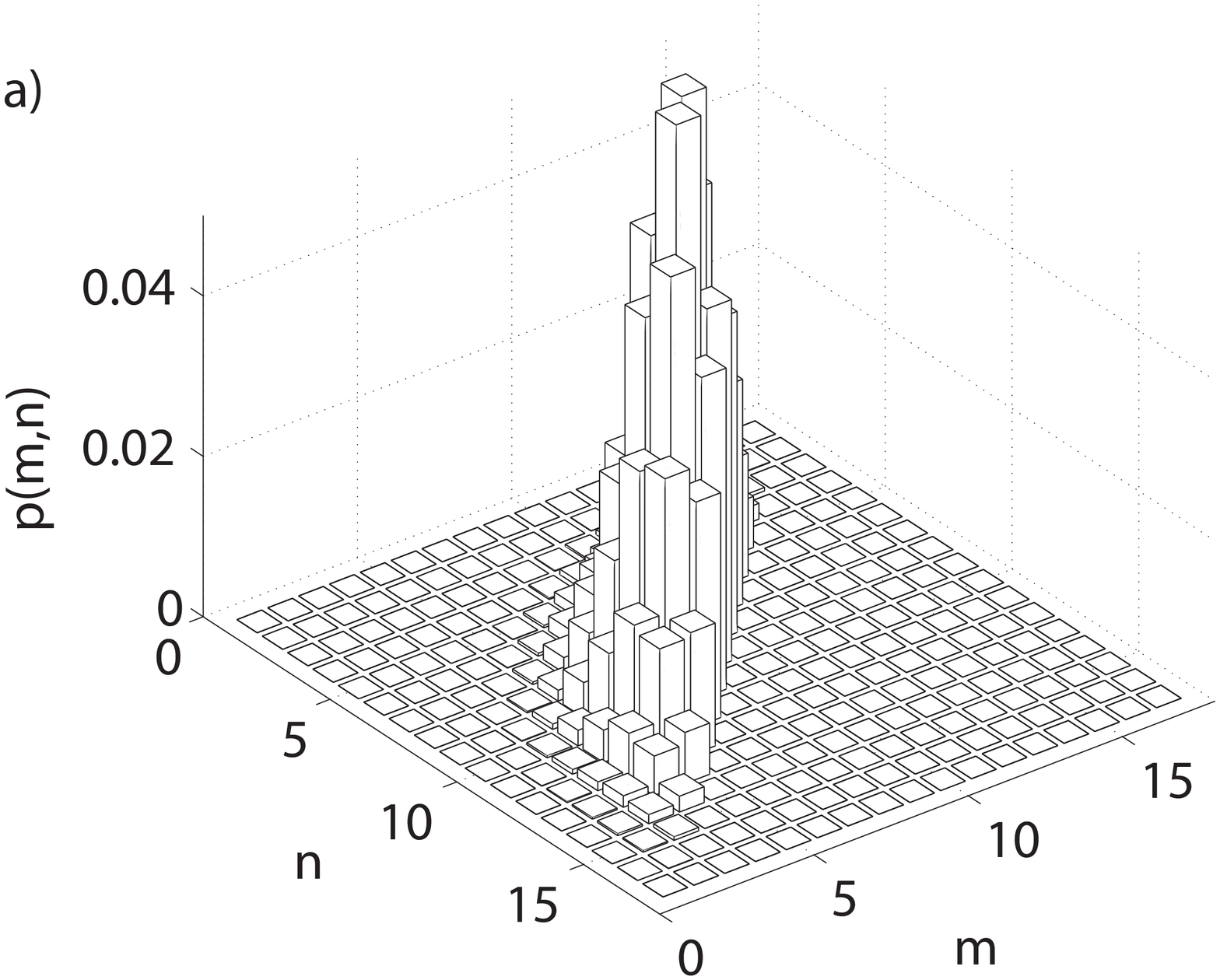,width=0.48\linewidth}  &\epsfig{file=
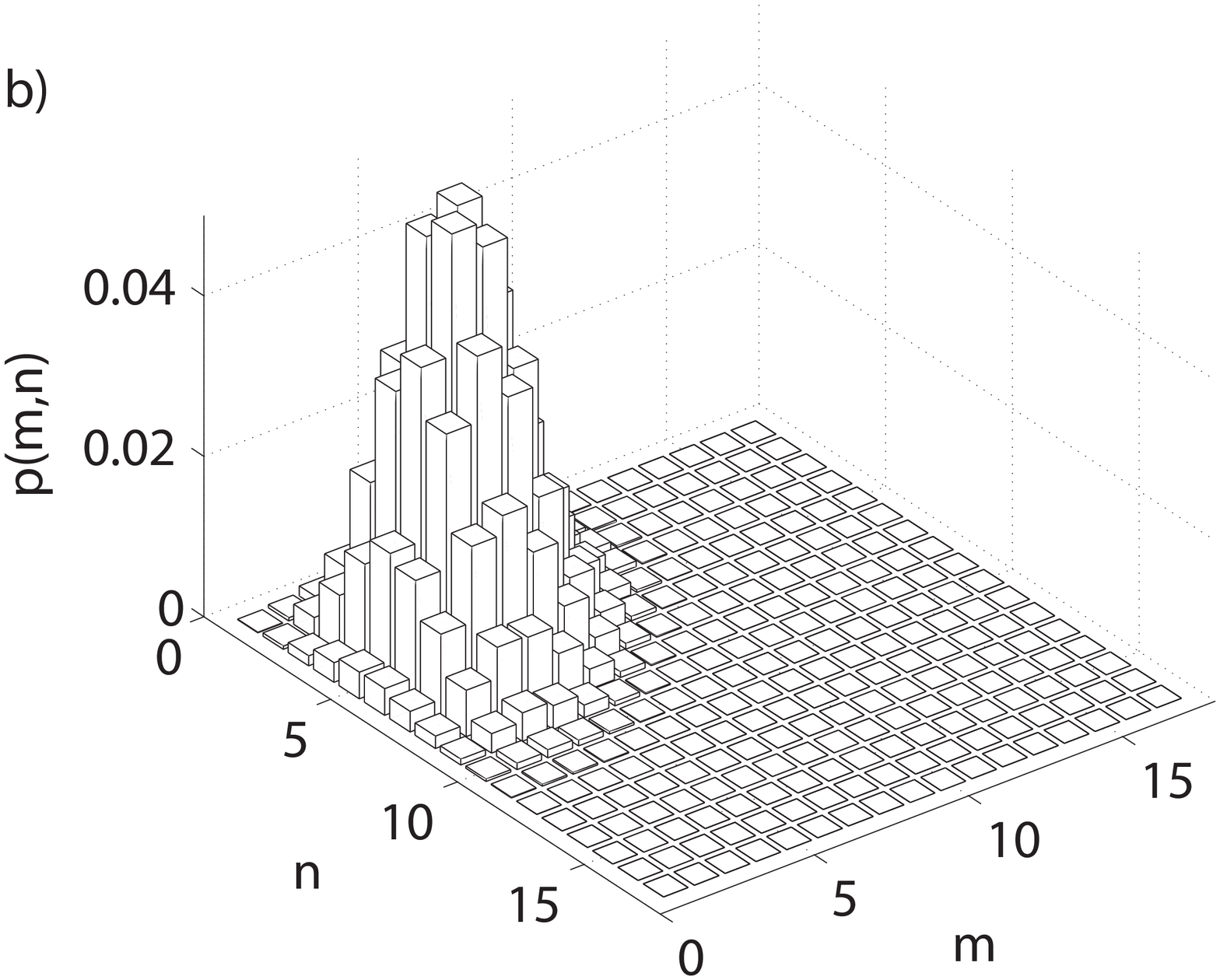,width=0.48\linewidth} \\
\epsfig{file= 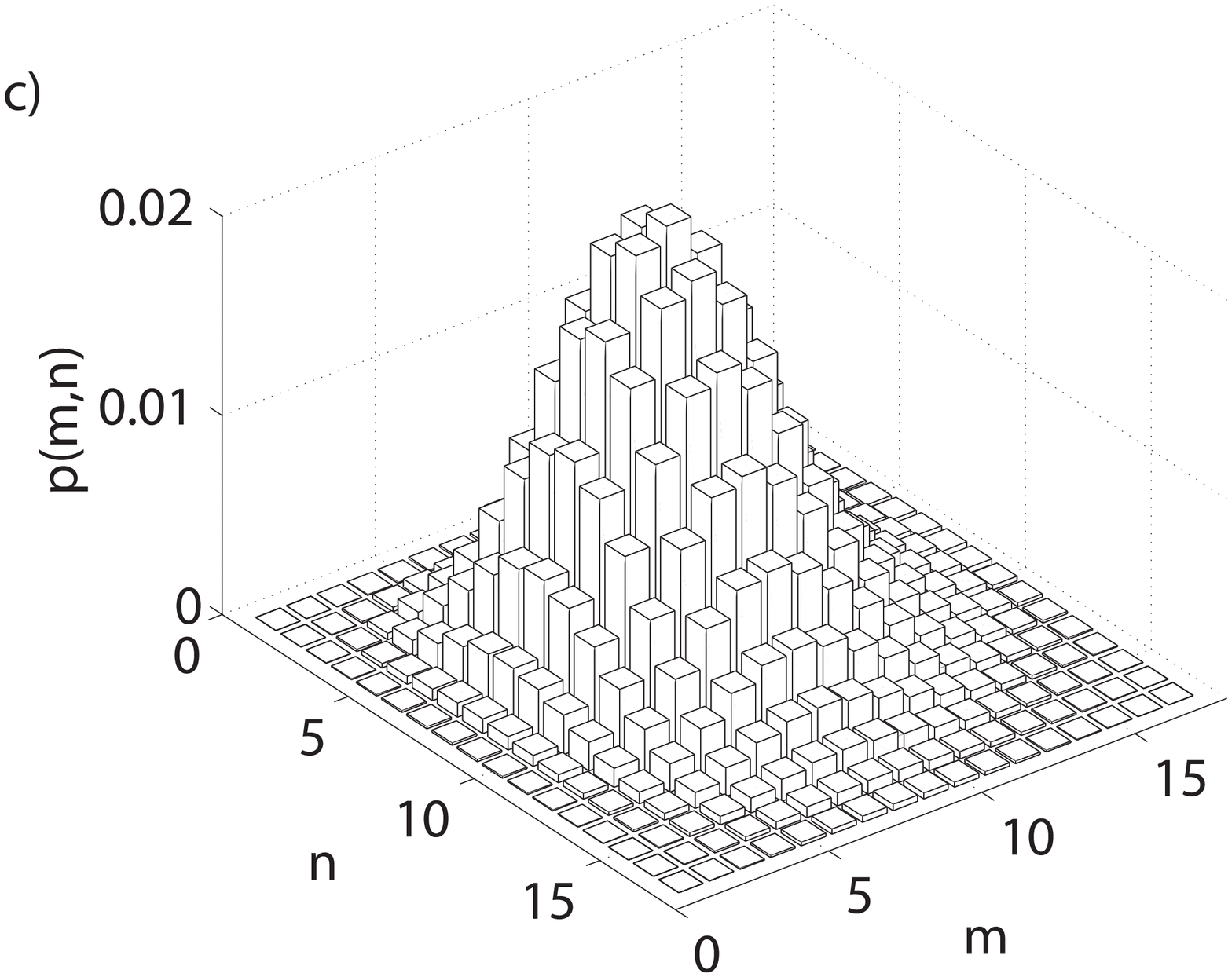,width=0.48\linewidth} & \epsfig{file=
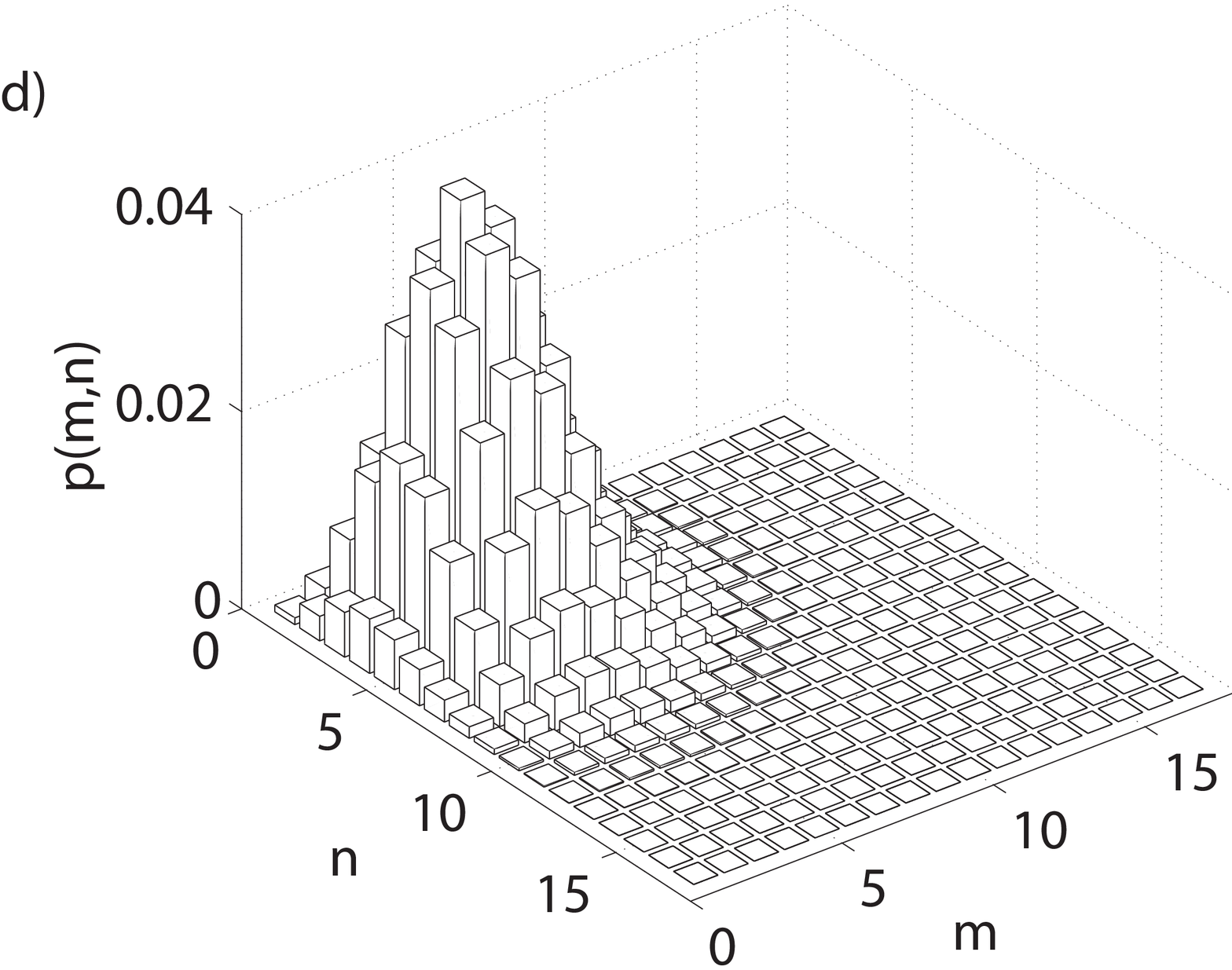,width=0.48\linewidth}
\end{tabular}
\caption{Joint probability distribution of an expanded MI (upper
row) and a SF (lower row) in a 4x4 lattice with two symmetrically
placed detectors. In fig. a) and c) $x_d=0$. In fig. b) and d)
$x_d=1$ cm. Parameters used: $z_0=1 cm$, $\Delta_z=2$ mm,
$\Delta_x=\Delta_y=2$ cm, $\kappa=0.5$.} \label{2det_mi_sf}
\end{figure}

\begin{figure}
\epsfig{file= 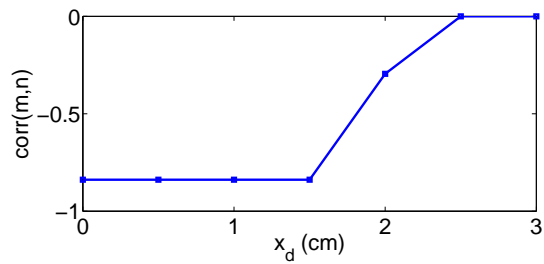,width=0.9\linewidth}
\caption{Correlations of the joint probability distribution for an
expanded MI state with two symmetrically placed detectors. As the distance between the detectors increases, the counting events at the two detectors are no longer correlated.
Parameters used: $z_0=1$ cm, $\Delta_z=2$ mm,
$\Delta_x=\Delta_y=2$ cm, $\kappa=0.5$} \label{fig-correlations}
\end{figure}

Let us now consider two detectors of the same size that are placed
symmetrically at a distance $\textbf{x}_1=(x_d,0,z_0)$ and
$\textbf{x}_2=(-x_d,0,z_0)$ from the lattice center. In the limit
of large detectors, we study the joint counting distribution of
the SF and MI state for different distances between the detectors.
Fig. \ref{2det_mi_sf} shows the counting distributions for two
overlapping detectors (left column) and for two detectors
separated by $2 x_d= 1$cm (right column). For the SF state, shown
in the lower row in figs. \ref{2det_mi_sf}, the joint counting
distribution is a Gaussian function for both cases. This is
expected, as the joint counting distribution eq. (\ref{eq-p_SF})
is a product of the single detector counting distributions. This
is analogous to the detection of coherent states of light. For the
MI state shown in the upper row in fig. \ref{2det_mi_sf}, we
observe a squeezed distribution, indicating the correlations of
the atoms counted at the two detectors. Note that as the distance
between the detectors increases, the squeezing of the distribution
is less pronounced. The correlations between the counting events
at the two detectors can be seen more clearly when looking at the
correlation function eq. (\ref{eq-correlations}). Note that for
the superfluid state, there is no difference between the joint
counting distribution and the product of the single particle
distributions. For the Mott state, we study the correlations for
varying distance between the two detectors $x_d$. In Fig.
\ref{fig-correlations}, we show how the correlations decrease when
increasing the distance between detectors $x_d$. Note that the
distance $x_d$ denotes the distance between the center of the two
detectors. For $x_d=0$, the detectors fully overlap, and for
$x_d>\Delta$ the detectors are completely separated.

\subsection{Detection of insulating states with different occupation patterns}\label{sec-checkboard}
Let us now focus on the detection of insulating states with
different occupation patterns by particle counting. As discussed
above, in order to detect the different patterns, the
crossed-correlations have to be of the order of the
autocorrelations. This is clear as away from the lattice, all the
on-site correlation terms become equal. Let us discuss the example
of a checkerboard state, where every second site is occupied, and
a state with stripes, where every second line is occupied. For the
striped state, the leading crossed-correlation terms eq.
(\ref{eq-off-diag-A}) are the ones that correspond to the nearest
neighbors. For the checkerboard state, where neighboring sites are
not occupied, the leading terms are the ones that correspond to
diagonally adjacent sites. In order to distinguish the different
patterns, it is thus essential that these two leading
crossed-terms are sufficiently different and at the same time
comparable to the on-site correlations. From Fig.
\ref{fig-Elements-A}, we see that this implies that the limit of
small detectors has to be considered. However, if the detector is
very small, all the terms are equal and the patterns are not
distinguishable. One should thus consider an intermediate detector
size.

\begin{figure}
\epsfig{file= 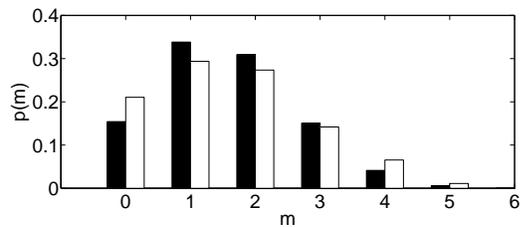,width=0.9\linewidth}
\caption{The counting distributions of an expanded one-dimensional checkerboard  (black bars) and striped insulating pattern
(white bars) are clearly distinguishable. Parameters used: $z_0=1$ cm, $\Delta=0.1$ cm, $\Delta_z=0.02$ mm,$\kappa=1$, $N_p=12$
.} \label{fig-checkerboard}
\end{figure}

In Fig. \ref{fig-checkerboard}, we illustrate the effect for a 1D
system of $N=12$ particles. We compare the counting distributions
of a checkerboard-like state, where every second site is occupied,
and a state where a block of six sites is occupied and a block of
six sites is empty. In order to distinguish the two states, from
Fig. \ref{fig-Elements-A}, we choose a detector size of
$\Delta=0.02$ mm, such that the ratio of the crossed-correlation
terms between neighboring sites and the autocorrelations is $0.6$.
Fig. \ref{fig-checkerboard} shows that the different occupation
patterns are reflected in the counting distribution.

\subsection{Detection of a Supersolid state}
\begin{figure}
\epsfig{file= 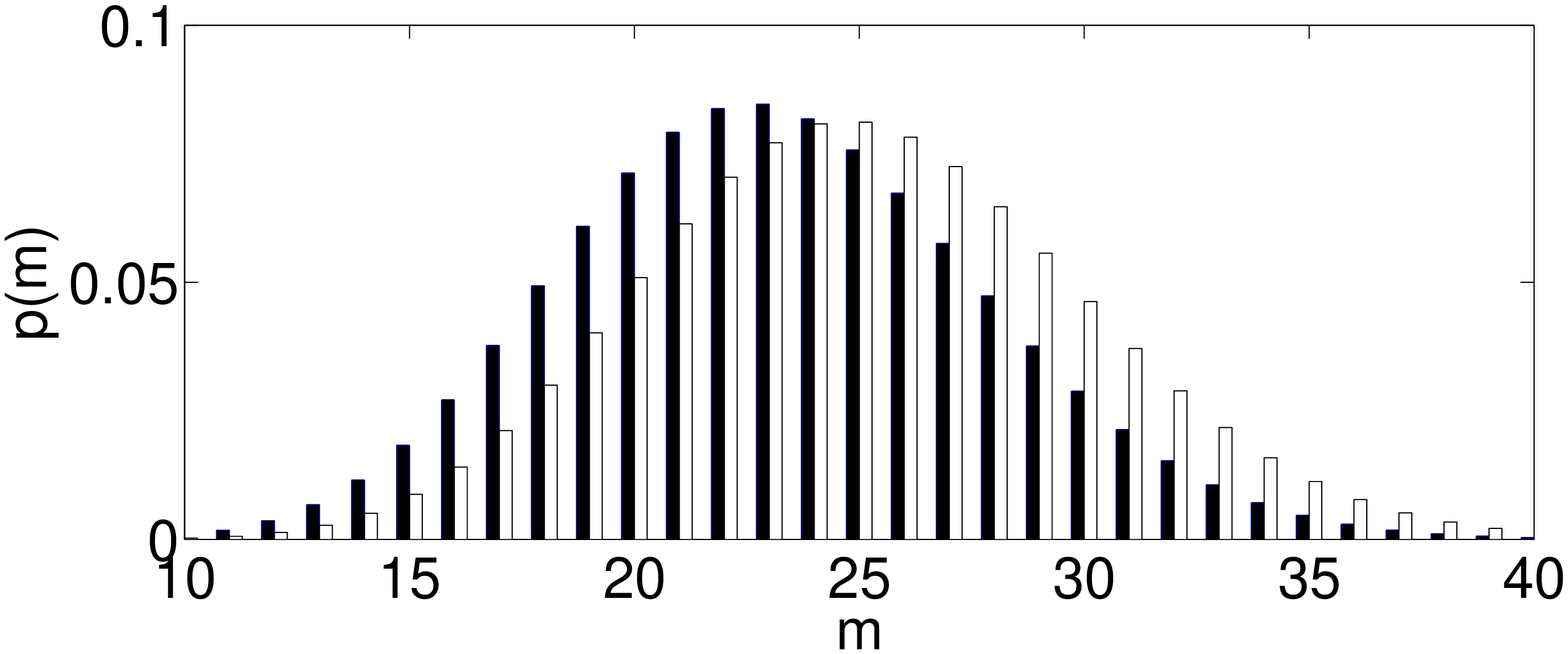,width=0.9\linewidth} \caption{The counting
distributions of an expanded supersolid state with $|\beta|^2=0.5$ and
$|\gamma|^2=1.5$ (black bars) and a superfluid state $|\alpha|^2=1$
(white bars) are clearly distinguishable. Parameters used: $z_0=1$ cm, $\Delta=1$ cm, $\Delta_z=0.02$ mm,
$\kappa=1$} \label{fig-SS-SF}
\end{figure}

As for the detection of states with different occupation patterns
in the insulating regime, the detection of supersolid states
\cite{Legget1970,Prokof2007} requires the limit where the
crossed-correlation terms for neighboring sites are comparable to
the auto-correlation terms. We consider a supersolid state with
$N$ sites and mean density $\alpha_{2i}=\beta$ and
$\alpha_{2i-1}=\gamma$. For the limit where the
crossed-correlation terms for neighboring sites are the only
non-negligible interference terms, the counting distribution eq.
(\ref{eq-p_SF}) is given by a Poissonian distribution with mean
\begin{eqnarray} \label{eq-p-SS} &&
\bar{m}=\frac{N}{2}A_d(\beta^2+\gamma^2)+2 NA_{NN}\beta\gamma,
\end{eqnarray}
where $A_d$ denotes the diagonal elements corresponding to the
on-site correlations and $A_{NN}$ denotes the nearest neighbor
crossed-correlation terms. Let us compare this to a superfluid
state with a homogeneous density per site,
$|\alpha_i|^2=\frac{|\beta|^2+|\gamma^2|}{2}$ for all $i$. The
counting distribution eq. (\ref{eq-p_SF}) is thus given by a
Poissonian distribution with mean
\begin{eqnarray}
\label{eq-p-SF2} \bar{m}=
\frac{N}{2}A_d(\beta^2+\gamma^2)+NA_{NN}(\beta^2+\gamma^2).
\end{eqnarray}
From eqs. (\ref{eq-p-SS}) and (\ref{eq-p-SF2}) it is clear that a
supersolid state can be distinguished from a superfluid state by
particle counting. In Fig. \ref{fig-SS-SF} we illustrate this by
comparing a supersolid state to a superfluid state.
\section{Summary}
We have studied the counting distributions of atoms falling from
an optical lattice and propagating in the gravitational field. The
intensity of atoms recorded at a detector located far away from an
optical lattice can be decomposed into autocorrelation and
crossed-correlations between the expanding modes. The ratio
between these terms depends crucially on the geometry of the
detector. In the limit when the detector is large compared to the
expanded modes, the crossed-correlation terms are negligible and
only long-range correlations of different states can be
distinguished. In this limit a SF state has a poissonian number
distribution while a MI has subpoissonian number distribution for
a detector of finite size located at a distance $z_0$ from the
lattice. The two states can also be readily distinguished from the
joint probability distribution of counting the particles at two
detectors. In the SF regime, the joint probability distribution is
a product of the two independent number distributions while in the
MI regime, the distributions are highly correlated.

When the detector is small compared to the expanded wave function,
the crossed-correlation terms for adjacent sites are of the order
of the auto-correlations. We have shown that by choosing the size
of the detector in an appropriate way, different occupation
patterns can be distinguished by particle counting after expansion
both in the insulating as well as in the superfluid regime.
\begin{acknowledgments}
We acknowledge financial support from the Spanish MICINN project
FIS2008-00784 (TOQATA), FIS2010-18799, Consolider Ingenio 2010
QOIT, EU-IP Project AQUTE, EU STREP project NAMEQUAM, ERC Advanced
Grant QUAGATUA, the Ministry of Education of the Generalitat de
Catalunya, and from the Humboldt Foundation. M.R. is grateful to
the MICINN of Spain for a Ram\'on y Cajal grant, M.L. acknowledges
the Alexander von Humboldt Foundation and Hamburg Theoretical
Physics Prize.
\end{acknowledgments}

\end{document}